# Ab initio molecular and solid state studies of $Fe^{II}$ spin cross-over system $[Fe(btz)_2(NCS)_2]$

L. Kabalan, S. F. Matar\*, M. Zakhour (\*\*) and J.F. Létard.

October 24, 2018

ICMCB, CNRS, Université Bordeaux 1. 87 Avenue du Dr. Albert Schweitzer, F-33608 Pessac Cedex, France
(\*\*) LCPM, Université Libanaise. Fanar-Beyrouth. Lebanon.

\*corresponding author: matar@icmcb-bordeaux.cnrs.fr

**Abstract**

Ab initio computations within the density functional theory are reported for the spin cross-over complex, [Fe(btz)$_2$(NCS)$_2$], where 3d$^6$ Fe$^{II}$ is characterized by High Spin (HS t$_{2g}^4$, e$_g^2$) and Low Spin (LS t$_{2g}^6$, e$_g^0$) states. Results of infra red and Raman spectra for the isolated molecule are complemented for the crystalline solid with a full account of the electronic band structure properties: the density of states assessing the crystal field effects and the chemical bonding assigning a specific role to the Fe-N interaction within the coordination sphere of Fe$^{II}$.

# 1 Introduction

Few inorganic transition metal ion complexes have the possibility of exhibiting two electronic states of d electrons, the High Spin (HS) and Low Spin (LS). Switching between these two states is subjected to small energy magnitudes around $\sim$ k$_{BT}$ and the transitions can be achieved with external constraints such as temperature, pressure as well as by applying light. Such spin cross-over (SCO) behaviour has been discussed in detail in several reviews [1, 2, 3, 4]. Theoretical studies at the molecular level [5, 6] are equally available but to our knowledge there are no solid state investigations of the electronic structure for the whole crystal system. Such studies are likely to provide further information on the changes of the electronic density of states as well as on the chemical bonding in the close neighbourhood of central Fe ion. The purpose of this work is to provide spectroscopic and electronic structure information on SCO system [Fe(btz)$_2$(NCS)$_2$] (*btz* stands for 2.2'-Bi-4.5-dihydrothiazine), both for the isolated molecule and for the extended solid state. The *btz* ligand is similar in its structure to 2,2'-bipyridine-N,N' aromatic ligand. However the presence of sulfur replacing carbon provides a non-planar bidentate ligand as it is shown by the projection in fig. 1. Lastly $NCS^-$ mono-dentate ligands are in a cis-position like in most of SCO complexes, such as $[Fe(phen)_2(NCS)_2]$ [7].

The crystallographic and SCO characteristics were investigated by many authors [8, 9, 10]. Just like the prototype of SCO systems, $[Fe(phen)_2(NCS)_2]$, this molecule keeps the orthorhombic space group (P*bcn*, 60) during the spin transition, the cell parameters remaining close and the molecular packing the same. In the case of $[Fe(btz)_2(NCS)_2]$, the SCO transition is not accompanied by a structural transition [11]. Although molecular arrangement in each of the two complexes is comparable: layers parallel with the $a, b$ plane (xOy basal plane). The intermolecular interactions at ambient temperature and their reinforcement at low temperature seem primarily of intra-layer type in $[Fe(phen)_2(NCS)_2]$ and inter-layer in $[Fe(btz)_2(NCS)_2]$. This underlines a coupling of different nature between close molecules which could be at the origin of cooperative effects [11]. Contrary to $[Fe(phen)_2(NCS)_2]$, the SCO transition in $[Fe(btz)_2(NCS)_2]$ is gradual and $T_{\frac{1}{2}}$ is about 215 K [12].

# 2 Theoretical and computational frameworks

Although the use of Hartree-Fock (HF) approach has been shown to provide a good description of the molecular orbital and chemical bonding properties mainly in organic chemistry, it becomes well established that calling for the density functional theory (DFT) framework [13] brings far more accurate results regarding the energetics and related properties. This is because the compulsory exchange and correlation (XC) effects are equally treated, albeit at a local level, within DFT while only exchange is well accounted for in HF although in a better way (exact exchange [14]) than in DFT. Taking the best out of each one of the two approaches led to improvements in ab initio molecular calculations with the so called "hybrid functionals". They consist of mixing exact HF exchange, ex. following Becke [14] and DFT based correlation, ex. following Lee, Yang and Parr, i.e., the so-called LYP correlation [15], with proportions that help to reproduce molecular properties of several systems. The calculations have been performed using $Gaussian03$ package [16].

In the solid state the molecule is built within the framework of an extended solid, according to the X-ray diffraction crystal structure determination [8, 9]. This allows for a better account of possible cooperative effects and for a description of the electronic band structure of the whole system. For this purpose we use the all electron augmented spherical wave method (ASW) [17]. Beside its use of DFT, ASW method is based on the atomic sphere approximation (ASA), a special form of muffin-tin approximation which consists in dividing the cell volume into atomic spheres whose total volume is equal to the cell volume. The calculation being carried out in the atomic spheres space, empty spheres (pseudo-atoms) need to be introduced -without symmetry breaking- in open (low compactness) structures such as that of the system studied here. Further, empty spheres allow for the iono-covalent characteristcis of the system to be accounted for by receiving charges from actual atomic species. The use of this method in molecular systems was formerly validated by Eyert et al. [18]. In the context of this study, ASW-LDA was used in order to obtain a description of the electronic structure with the partial, site projected, density of states. Further the chemical interaction is assessed especially within the coordination sphere of divalent iron thanks to $E_{COV}$ (covalence energy), which allows to get two-body chemical bond characteristics [19]. Negative, positive and zero $E_{COV}$ point to bonding, anti-bonding and non-bonding interactions respectively.

# 3 Results and discussion for the isolated molecule

Starting geometries for HS and LS configuration molecules were extracted from the Crystallographic Information File (CIF) files [8]. After testing different functionals and basis sets through preliminary trial computations, we found most accurate results for the energy and vibrational spectra in LS and HS states with hybrid functional B3LYP and large and complete enough basis sets such as so-called $6-31G$ and $6-311G$, as well as $LanL2DZ$ basis sets. The latter include double-$\zeta$ with the Los Alamos effective core potential for Fe, S and the Dunning-Huzinaga all-electron double-$\zeta$ basis set with polarization functions for the H, C, and N atoms ($\zeta$ is the exponent in the Gaussian type orbitals GTO) [20, 21]. Firstly the geometry of the molecule is optimized in both spin states and the results tested against the experimental distances within the coordination sphere of iron (cf. fig. 1). Table 1 gives the obtained distances and angles for LS and HS configuration from B3LYP/$LanL2DZ$ calculations; we note however that close values were equally obtained with 6-311G basis set. Although the final Fe-N distances do not translate a perfect octahedral O$_h$-like environment, we find in the average d$_{LS}(Fe-N) \sim 1.990$ Å, smaller than d$_{HS}(Fe-N) \sim 2.186$ Å. This is an expected result due to the increase of the volume of divalent Fe coordination sphere arising from the occupation of anti-bonding e$_g^*$ states with two unpaired electrons: LS ($t_{2g}^6$ $e_g^0$) HS($t_{2g}^4$ $e_g^2$). These average values agree with theoretical ones for other similar SCO complexes in the literature [6] where BP86 hybrid functional and triple-$\zeta$ basis set were used. Similarly, an agreement between computed values of distances versus experiment was also found. However some discrepancy was observed for the magnitudes of $Fe - \widehat{N - C}$ and $(SCN) - \widehat{Fe - (NCS)}$ angles, not only with $LanL2DZ$ basis set but with all other basis sets used. From the geometry optimized configurations with B3LYP/$LanL2DZ$, the infra-red (IR) and Raman theoretical frequency spectra were computed for the purpose of providing a spectroscopic signature of the system in its two spin states. The differences that can be observed for the positions as well as for the relative intensities between IR and Raman spectra in both HS and LS states point to the fact that there are active modes in IR which are not so in Raman and *vice versa* (cf. figs. 2 *versus* 3).

From this we have spectrocopic signatures for the systems not only at the level of the spin states (LS/HS) but for the IR/Raman spectroscopies as well. For instance this should be particularly helpful when micro-Raman charaterizations are carried out. Firstly we discuss the domain of low frequencies in wave numbers: cm$^{-1}$. At $\nu < 500$ cm$^{-1}$ the Fe-N elongation modes are present. In the case of LS, a peak at 532.1 cm$^{-1}$ could be detected and assigned to Fe-N$_{(NCS)}$ elongation which is mainly active in IR. This magnitude comes close to experiment; i.e., 532.6 cm$^{-1}$ [3, 6]). For the HS case this same elongation is found at $\nu$ =240.7 cm$^{-1}$ (exp. 249.0 cm$^{-1}$[3, 6]). Fe-N$_{(btz)}$ HS/LS elongation values are 221/367 cm$^{-1}$; they follow the same trend as above for the magnitudes however we have no experimental data available to confront with. The other relevant frequencies characterizing the HS/LS varieties which could serve as a signature, are those of $NCS^-$ ligand; their magnitudes and attributions are given together with experimental results in table 2 with a good agreement. Values for LS state are systematically higher than for HS which can be understood if one considers elementary citeria such as the average distances which are shorter for the LS complex (cf. table 1). The frequency of the electromagnetic wave which induces the vibration of elongation is given by the relation $\nu=\frac{1}{2\pi}\sqrt{\frac{k}{\mu}}$ where k is the constant of bonding strength (considered here as a spring), proportional to the binding energy and $\mu$ the reduced mass of the two atoms connected by this bond. The distances in the case of LS are smaller than in HS case; thus the ratio k($\frac{HS}{LS}$) = 0.962, has a magnitude lower than 1.

At this level of development of the results, it becomes relevant to discuss the energetics, $\Delta_{LS/HS}$, between the two spin states with the functional and basis set utilized. The average value sobtained for $\Delta_{HS-LS}$ is $\sim$ -2.5 kJ/mol favoring the high spin case. We note that the $\Delta_{LS/HS}$ is governed by the amount of exchange admixture in the energy functional [22, 23]. A main result was that adjustment of the exact exchange admixture is necessary since pure density functionals tend to favor low-spin states, while hybrid functionals like B3LYP usually give the high-spin state as the lowest-energy state if the low-spin/high-spin energy splitting $\Delta_{LS/HS}$ is not too large [24]. Our results are in agreement with those of Paulsen, Trautwein and co-works [5] who have conducted an extensive study on several spin-crossover complexes using different basis sets and density functionals. Their results show that the B3LYP favors, in general, the high-spin state while all pure density functionals favor the low-spin state. Interestingly we shall show that the energetics extracted from the solid state calculations (section 4) will exhibit this tendency.

Further we have tried to reproduce the T$\frac{1}{2}$ characteristic temperature from the basis sets that were used. At the transition which is mainly driven by vibration energies, equal proportions of HS and LS populations are co-existing and one defines T$\frac{1}{2}$ from $\Delta G = \Delta H - T\Delta S$; where G, H and S are the well known letters indicating free energy, enthalpy and entropy. At this equilibrium $\Delta G = 0$, and T leads to T$\frac{1}{2}$ defined as $\frac{\Delta H}{\Delta S}$. The electronic contribution to the entropy arises from the HS configuration since $\Delta S_{el} =$ R(Ln(2s+1)$_{HS}$ -Ln(2s+1)$_{LS}$; we signal total spins by script s in order to differenciate from entropy S; further R = 8.3 J [6]. Since all spins are paired in the t$_{2g}$ manifold, t$_{2g}^6$, s= 0 while it s= 2 for HS. Then S$_{tot}$ = S$_{HS}$ =

R Ln(2 s$_{HS}$+1) = 13.4 J. mol$^{-1}K^{-1}$. The computed value we obtain is 13.38 J. mol$^{-1}K^{-1}$ which comes very close to the formal one. Adding S$_{el}$ to S$_{vibrational}$ leads to compute T$\frac{1}{2}$ $\sim$ 100K which is about half the experimental one (215K). Such a discrepancy can be well obtained with calculations on the single molecule such as [Fe(phen)$_2$(NCS)$_2$] for which a magnitude of 1530K (T$\frac{1}{2}$(exp.)=176 K [7]) was obtained in the literature with BP86 hybrid functional [22].

One can interpret the reduced T$\frac{1}{2}$ value from the fact that we have a calculated $\Delta H$ = 6.4 kJ/mol while the magnitude we extract from the experimental magnetic susceptibility, $\chi_M$ $T = f(T)$ curve [11], amounts to 24 kJ/mol. Further $\Delta S$ is smaller than that obtained by the thermodynamic relation $\Delta G = \Delta H$ -T$\Delta S$ which gives $\Delta S$= 112 joules while it is 66 joules whith the 6-31g basis-set calculations. This raises modeling problems in reproducing experimental magnitudes of T$\frac{1}{2}$ which can have several origins, one being the cooperative effects which call for the use of neighbouring adjacent molecular species beside the single isolated molecule. We expect improvements of T$\frac{1}{2}$ provided such neighbouring molecules be adjoined to the complex. The calculations are underway.

# 4 Calculations for the extended solid

To gain insight into the role of each atomic species in the electronic band structure and the chemical bonding, we carried out all electrons, spin degenerate, calculations for the tetra-molecular [Fe(btz)$_2$(NCS)$_2$] system in the orthorhombic P*bcn* experimental structural parameters [8, 9]. Our calculations show large charge transfer, translating ionic behaviour, from the atomic species Fe, N, C and S to the empty spheres (ES). Remember that ES are pseudo-atoms with zero-core, introduced as stated above, for an account of the large voids within the open structure of the molecular system. The plots of the projected density of states (PDOS) for the constituent atom sites are presented in figs. 4 and 5 for LS and HS configurations respectively. Along the $x$-axis, the energies are considered with respect to the top of the valence band VB (E$_V$), not the Fermi energy (E$_F$), because the system is close to an insulator, i.e. there is a DOS minimum at this energy boundary, although the gap can be hardly seen. This drawback can be attributed to the use of plain LDA functional based on the homogeneous electron gas. It is worth saying here that the insulating character is likelier to be obtained if LDA+U heavier calculations were carried out; U is the on-site Coulomb repulsion parameter which enhances the correlation. However the purpose of our work is to understand the different behaviours of our system in its two spin states. This is done by discussing the PDOS which are most significant in the neighbourhood of $E_V$. For this reason we consider a narrow energy window, -6,6 eV range. There can be seen a dominant presence of Fe PDOS (red solid lines) on both sides of $E_V$. Its chemical bonding with other constituents species, especially N, is observed in the lower part of the VB, see next section. The original interesting feature which can be traced out is relevant to the intrinsic properties of the complex system in LS and HS states. The major different feature between figs. 4 and 5 is relevant to the splitting of Fe(d) states according to the crystal field expected for LS and HS, i.e. the empty $e_g$-like manifolds PDOS are high in energy within the conduction band (CB) (fig. 4) in LS with a separation of ∼2.5 eV or 242 kJ/mol. which comes close to typical crystal field splitting magnitude $\Delta_0$ in octahedral-like environments. On the contrary the partially filled HS $e_g$-like manifold is close to $E_V$ (fig. 5).

The variational energy difference between HS and LS ASW-LDA calculations carried out with the same Brillouin zone **k** mesh precision amounts to $\Delta E_{LS-HS}$ = -8374.6 J for 4 formula units (fu), i.e., -2.09 kJ/fu. While this favors the LS configuration energetically, contrary to all molecular calculations above, it is interesting to note a close magnitude to the average value obtained from single molecule calculations.

Regarding the chemical bonding within the complex system one mainly needs to examine the Fe-N bond. Considering the LS configuration (the same reasoning holds for the HS one), table 1 gives the shorter $d_{Fe-NCS}$=1.979 Å distances than $d_{Fe-N(btz)}$ (1.987 and 2.005 Å) so one expects relatively stronger bonding for the former. The ECOV plots shown in fig. 6 exhibit a mainly bonding (negative magnitude) Fe-N character within the VB. However a striking feature of a largely dominant Fe-$N_{NCS}$ interaction is mainly present. While one can assign a certain role to the smaller $d_{Fe-NCS}$ as with respect to the two other distances, leading to a stronger Fe-N bong between Fe and NCS ligand, one needs to say that the major bonding is due to the charged ligands only. This gives the following picture "$Fe^{2+}(NCS^-)_2$"$(btz)_2$, i.e. with the *btz* ligand playing a secondary role in the coordination sphere of $Fe^{2+}$.

# 5 Conclusion

In this work, we presented a molecular calculation for the [Fe(btz)$_2$(NCS)$_2$] complex. The change of metal-ligand bond lengths upon spin-flip from singlet to quintet which is rationalized in terms of a one-particle picture as an occupation of molecular orbitals with anti-bonding character, is well-reproduced in the calculations at the level of the single molecule. The calculated IR and Raman absorption spectra for the low-spin and high-spin, we obtained with the *LanL2DZ* frequency of antisymmetric and symmetric NCS- were very close to those obtained by experiments. We also carried out original calculations for the solid states accounting for the whole crystalline tetra-molecular complex. With the all-electrons ASW-DFT method we obtained the PDOS in the two states and the chemical bonding characteristics. The first show in the close neighbourhood of the top of the valence band the change from LS to HS behaviour within the Fe coordination sphere for the crystal field. The chemical bonding shows the original feature of the predominant $Fe^{2+}$-$NCS^-$ bond with respect to much lower magnitude Fe-N(btz) crown.

# 6 Acknowledgments

Lara Kabalan thanks the CNRS$L$, Lebanese Council of Research, for her Ph.D. scholarship. We acknowledge computational facilities provided by the University Bordeaux 1 within the M3PEC *Mesocentre Regional* (http://www.m3pec.u-bordeaux1.fr) supercomputers.

# References

[1] P. Gütlich, Y. Garcia, H. Goodwin, A. Chem. Soc. Rev., 29, 419-427 (2000).

[2] Gütlich and H.A. Goodwin, Spin Crossover in Transition Metal Compounds, Topics in Current Chemistry , Springer Wien, New-York, 233-235 (2004).

[3] P. Gütlich, structure and bonding, 44, 83-195 (1981).

[4] J. F. Létard, P. Guionneau and L. Goux-Capes, Topics in Current Chemistry, 235, 221-249, (2004).

[5] H. Paulsen, L. Duelund, H. Toftlund, A. Trautwein, Inorg.chem., 40, 2201-2203 (2001).

[6] C. Brehm, M. Reiher and S. Schneider, J. Phys. Chem. A, 106, 12024 (2002).

[7] B. Gallois, J.A. Real, C. Hauw and J. Zarembowitch. Inorg. Chem., 29, 1152 (1990).

[8] J. A. Real, B. Gallois, T.Granier, F. Suez-Panana and J. Zarembowitch, Inorg. Chem. 31, 4972-4979, (1992).

[9] T. Granier, B. Gallois, J. A. Real, J. Gaultier and J. Zarembowitch, Inorg. Chem. 32, 5305-5312 (1993).

[10] T. Granier, B. Gallois, J. A. Real, J. Gaultier and J. Zarembowitch, Inorg. Chem. 32, 221-249, (2004).

[11] F. Suez-Panama-Bouto, Ph.D. Thesis, Université Bordeaux 1 (1991).

[12] P.Guinneau, M. Marchivie, G. Bravic, J. F. Lé tard and Daniel Chasseau, Topics in Current Chemistry, 234, 97-128, (2004).

[13] P. Honenberg and W. Kohn, Phys. Rev. 136, B864 (1964); W. Kohn and L.J. Sham, Phys. Rev.140, A1133 (1965).

[14] V. D. Becke, J. Chem. Phys. 88 2547 (1988).

[15] C. Lee, W. Yang and R.G. Parr, Phys. Rev. B 37, 785 (1988).

[16] Gaussian 03, Revision C.02, M. J. Frisch, G. W. Trucks, H. B. Schlegel, G. E. Scuseria, M. A. Robb, J. R. Cheeseman, J. A. Montgomery, Jr., T. Vreven, K. N. Kudin, J. C. Burant, J. M. Millam, S. S. Iyengar, J. Tomasi, V. Barone, B. Mennucci, M. Cossi, G. Scalmani, N. Rega, G. A. Petersson, H. Nakatsuji, M. Hada, M. Ehara, K. Toyota, R. Fukuda, J. Hasegawa, M. Ishida, T. Nakajima, Y. Honda, O. Kitao, H. Nakai, M. Klene, X. Li, J. E. Knox, H. P. Hratchian, J. B. Cross, V. Bakken, C. Adamo, J. Jaramillo, R. Gomperts, R. E. Stratmann, O. Yazyev, A. J. Austin, R. Cammi, C. Pomelli, J. W. Ochterski, P. Y. Ayala, K. Morokuma, G. A. Voth, P. Salvador, J. J. Dannenberg, V. G. Zakrzewski, S. Dapprich, A. D. Daniels, M. C. Strain, O. Farkas, D. K. Malick, A. D. Rabuck, K. Raghavachari, J. B. Foresman, J. V. Ortiz, Q. Cui, A. G. Baboul, S. Clifford, J. Cioslowski, B. B. Stefanov, G. Liu, A. Liashenko, P. Piskorz, I. Komaromi, R. L. Martin, D. J. Fox, T. Keith, M. A. Al-Laham, C. Y. Peng, A. Nanayakkara, M. Challacombe, P. M. W. Gill,

B. Johnson, W. Chen, M. W. Wong, C. Gonzalez, and J. A. Pople, Gaussian, Inc., Wallingford CT, 2004.

[17] A. R. Williams, J. Kübler and C.D. Gelatt Jr. Phys. Rev. B **19** 6094 (1979); and V. Eyert, *The Augmented Spherical Wave Method – A Comprehensive Treatment, Lecture Notes in Physics* (Springer, Heidelberg, 2007).

[18] V. Eyert, B. Siberchicot, M. Verdaguer, Phys. Rev. B, 56, 14 (1997).

[19] G. Bester and M. Fähnle, J. Phys: Condens. Matter **13**, 11541 (2001).

[20] T. H. Dunning, Jr. and P. J. Hay, in Modern Theoretical Chemistry, edited by H. F. Shaefer III , 3, 1-28 (1976).

[21] P. J. Hay and W.R. Wadt, J. Chem.Phys., 82, 270 (1985).

[22] M. Reiher, O. Salomon, B. A. Hess, Theor. Chem. Phys.,107, 48-55 (2001).

[23] O. Salomon, M.Reiher, B. A.Hess, J. Chem. Phys.,117, 4729-4737 (2002).

[24] M.Reiher, Inorg. Chem, 41, 6928-6935 (2002).

|  | LS | LS exp.[8, 9] | HS | HS exp.[8, 9] |
|---|---|---|---|---|
| $d_{Fe-NCS}$ | 1.979 | 1.947 | 2.085 | 2.066 |
| $d_{Fe-N(btz)}$ | 1.987 | 1.967 | 2.267 | 2.177 |
| $d_{Fe-N(btz)}$ | 2.005 | 1.972 | 2.207 | 2.166 |
| $\alpha(Fe-N-C)^o$ | 155.20 | 169.1 | 145.30 | 167.80 |
| $\alpha(NCS)-Fe-(NCS)^o$ | 96.132 | 86.06 | 107.72 | 91.53 |
| $\alpha(N_{btz})-Fe-(N_{btz})^o$ | 80.0 | 80.0 | 73.92 | 74.7 |
| $\langle d_{Fe-N} \rangle$ | 1.970 | 1.979 | 2.186 | 2.136 |

Table 1: Distances in Å and angles compared to experiment obtained by using $LanL2DZ$ basis set and B3LYP hybrid functional for the molecule geometry optimisation in LS and HS [Fe(btz)$_2$(NCS)$_2$] complex. Average $\langle d_{Fe-N} \rangle$ distances are provided for sake of discussion (see text).

| $LS$ IR | $LS$ Raman | $HS$ IR | $HS$ Raman |
|---|---|---|---|
| 2103.89 (2100) | 2103.89 (2107) | 2070.71 (2060) | 2070.71 (2062) |
| 2113.51 (2110) | 2113.51 (2110) | 2077.78 (2070) | 2077.78 (2072) |

Table 2: Comparison of antisymmetric and symmetric N-C-S frequencies in cm$^{-1}$ between LS and HS [Fe(btz)$_2$(NCS)$_2$] complex. Experimental values from [3, 6] are given between brackets.

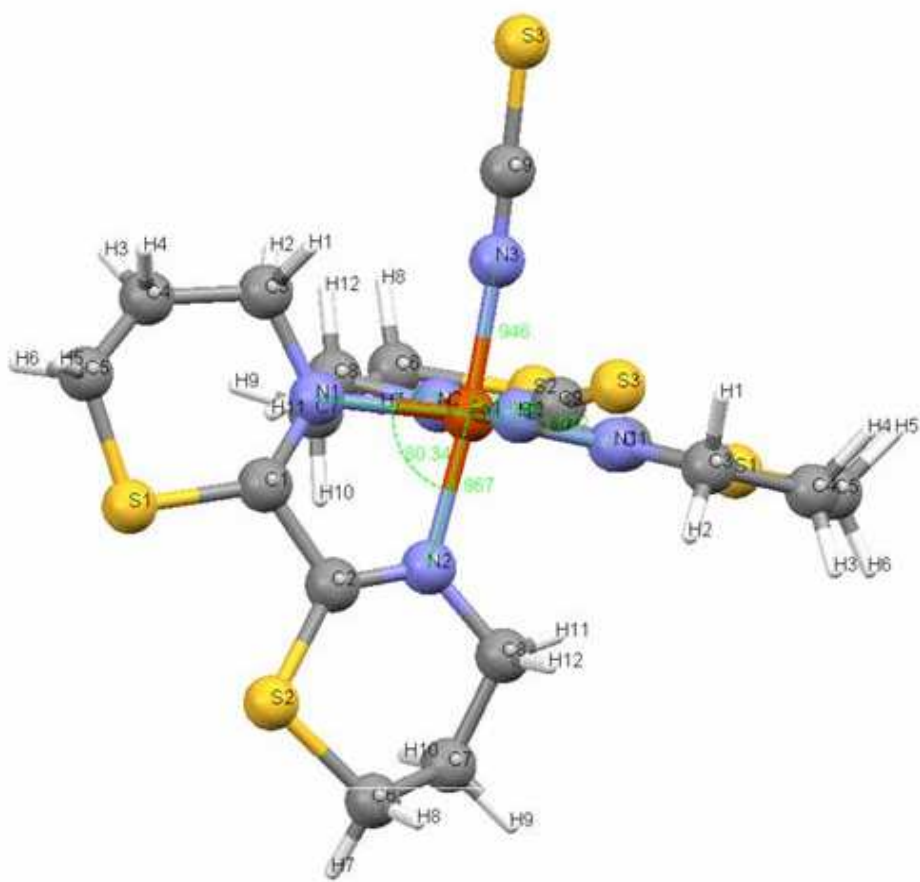

Figure 1: *(Color online).* Sketch of the molecular structure of SCO complex [Fe(btz)$_2$(NCS)$_2$] with characteritic distances (Å) and angles shown here for the low spin LS case.

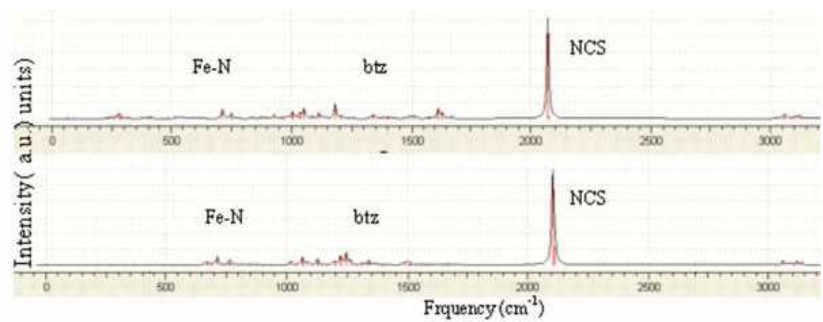

Figure 2: IR spectra of HS (in the top) and LS (in the down) of [Fe(btz)$_2$(NCS)$_2$] complex with the $LanL2DZ$ basis-set.

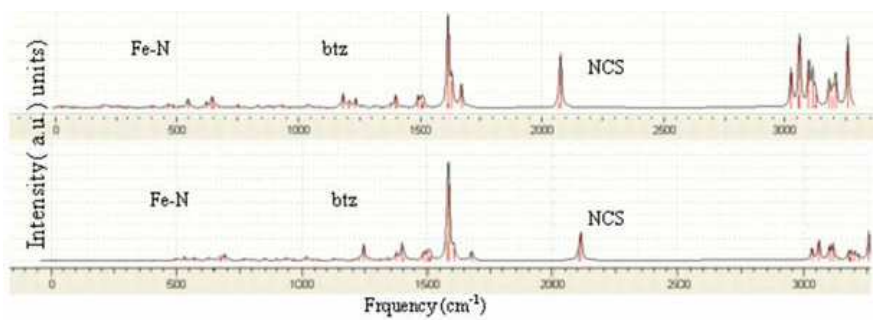

Figure 3: Raman spectra of HS (in the top) and LS (in the down) of [Fe(btz)$_2$(NCS)$_2$] complex with the $LanL2DZ$ basis-set.

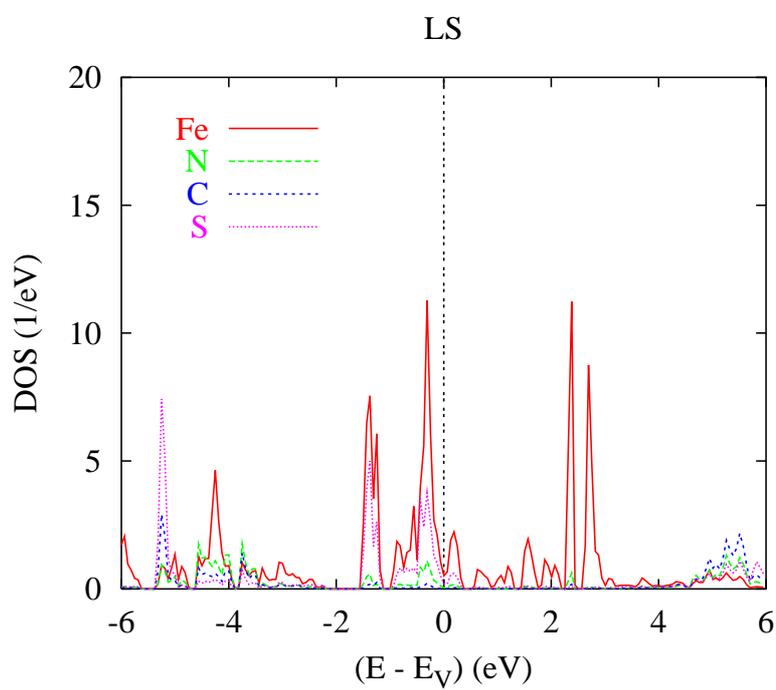

Figure 4: *(Color online)*. Site projected density of states (PDOS) of LS [Fe(btz)$_2$(NCS)$_2$]. Energy reference is with respect to the top of the valence band (E$_V$).

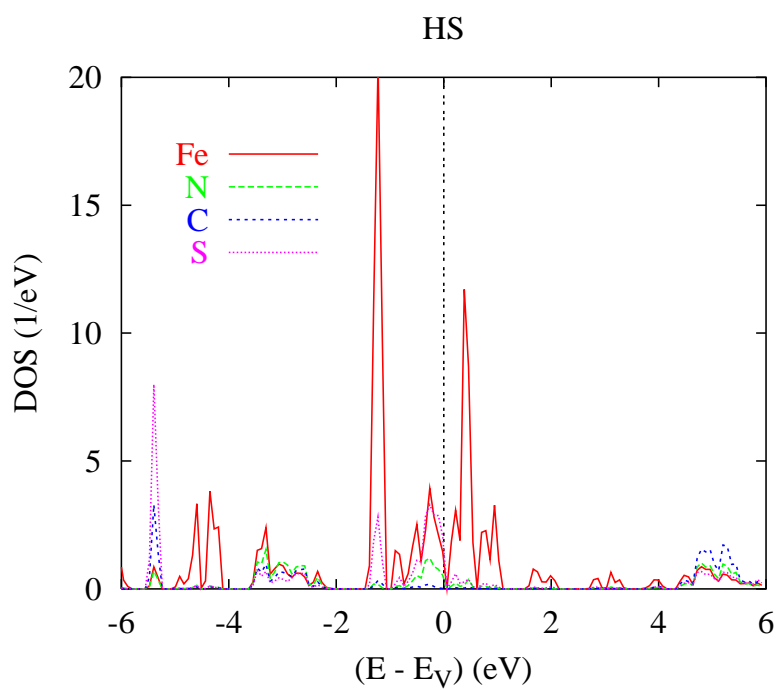

Figure 5: *(Color online).* Site projected density of states (PDOS) of [Fe(btz)$_2$(NCS)$_2$] in the high spin (HS) case. Energy reference is with respect to the top of the valence band (E$_V$).

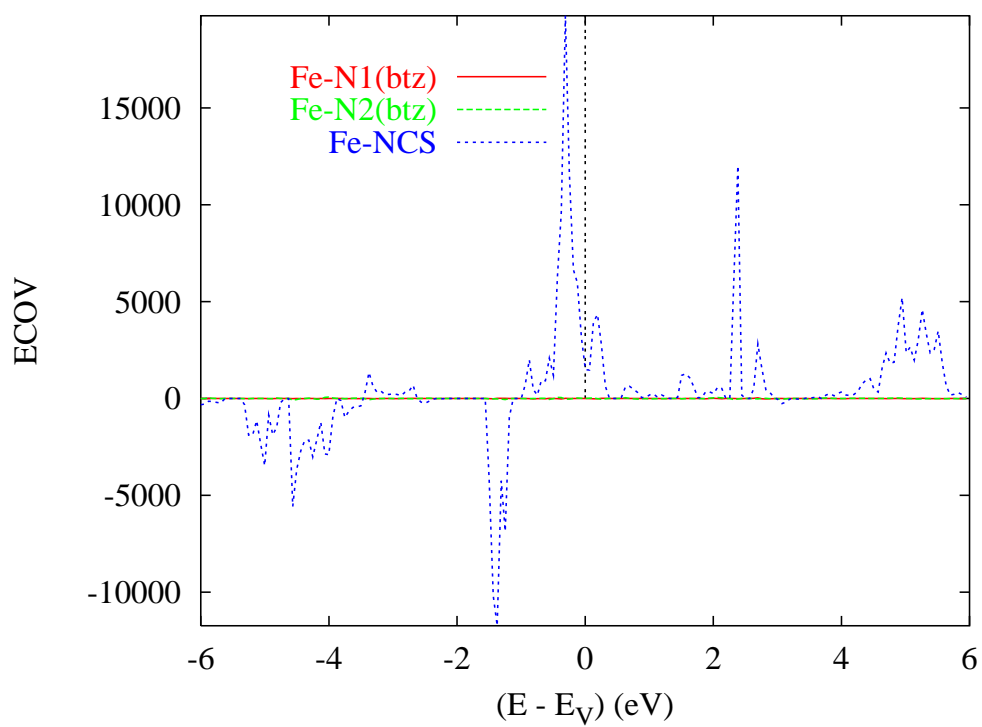

Figure 6: *(Color online)*. Chemical bonding following ECOV criterion (arbitrary units) for the different Fe-N interations within [Fe(btz)$_2$(NCS)$_2$] system in the low spin (LS) case. Energy reference is with respect to the top of the valence band (E$_V$).